\documentclass[11pt]{article}
\pdfoutput=1
\usepackage[export]{adjustbox}
\usepackage{jheppub}
\usepackage{bbm}
\usepackage{tikz}
\usetikzlibrary{calc}
\usepackage{graphicx,subcaption}
\usepackage{mathtools}
\usepackage{bm}
\usepackage{enumitem}
\usepackage{verbatim}
\usepackage{amssymb}
\usepackage{slashed}

\newcommand{\pisq}{{\pi^2}} 
\newcommand{\picu}{{\pi^3}} 

\newcommand{\ba}{{\bm\alpha}}
\newcommand{\blank}[1]{}
\newcommand{\NS}{{\text{NS}}}
\newcommand{\R}{{\text{R}}}

\newcommand{\hq}{{\hat q}}
\newcommand{\htau}{{\hat \tau}}

\newcommand{\cE}{{\cal E}}

\newcommand{\GW}[1]{{\color{orange}\textbf{#1}}}
\newcommand{\MD}[1]{{\color{green}\textbf{#1}}}
\renewcommand{\GW}[1]{{{#1}}}
\renewcommand{\MD}[1]{{{#1}}}

\newcommand{\bfmtb}{\bfmtb}
\renewcommand{\bfmtb}{\lambda}

\DeclareMathOperator{\Tr}{Tr}

\DeclareMathAlphabet{\mathbfsf}{OT1}{cmss}{bx}{n}

\newcommand{\N}{\mathbb{N}}
\newcommand{\Z}{\mathbb{Z}}

\newcommand\be{\begin{equation}}
\newcommand\ee{\end{equation}}
\newcommand\beq{\begin{equation}}
\newcommand\eeq{\end{equation}}
\newcommand\bea{\begin{eqnarray}}
\newcommand\eea{\end{eqnarray}}

\renewcommand{\a}{\alpha}
\renewcommand{\b}{\beta}

\renewcommand{\t}{\tau}

\newcommand{\rd}{{\rm d}}
\newcommand{\dx}{{{\rm d} x}}

\renewcommand{\tilde}{\widetilde}
\renewcommand{\hat}{\widehat}

\begin{document}

\title{Free fermions, KdV charges, generalised Gibbs ensembles, modular transforms and line defects}

\author{Max Downing}
\author{and G\'erard M.~T.~Watts}
\affiliation{Department of Mathematics,\\King's College London,\\
Strand, London WC2R 2LS, United Kingdom}
 \emailAdd{gerard.watts@kcl.ac.uk}
 \emailAdd{max.downing@kcl.ac.uk}
 
\abstract{
In this paper we return to the question of the modular properties of a generalised Gibbs ensemble of a single free fermion. We extend our previous proposals to a GGE containing an arbitrary number of conserved charges and provide a physical interpretation of the result in terms of a line defect. The defect description perfectly explains the product formula for the modular transformation we found previously. We also give a proposal for a Hamiltonian approach to the line defect.
}

\maketitle

\section{Introduction}

In this paper we return to the interpretation and generalisation of the results we obtained in \cite{Downing:2021mfw}. In that paper we proposed an exact formula for the modular transform of a generalised Gibbs ensemble (GGE) of a single massless free fermion in the presence of a single extra conserved quantity. 
In this paper we provide a physical interpretation for this formula (which was lacking in \cite{Downing:2021mfw}) and we extend the results to an arbitrary finite combination of conserved charges. {The physical interpretation is given by introducing a defect into the system. The presence of this defect can be seen in the TBA equations which were used in \cite{Downing:2021mfw} to derive our conjecture for the modular transform of the GGE \GW{and provides an explanation for the peculiar product form of the modular transform.} This defect can also be constructed explicitly in a Hamiltonian formalism.}

We also note that a specialisation of our results to the case $q=1$ and a single conserved quantity had already been proven in the mathematical literature \cite{hrj:8932}, and that a complete proof of these results for a finite collection of charges can be found in the companion paper \cite{downing2023modular}. For the background to and motivation for this work, we refer the reader to \cite{Downing:2021mfw}.

We start in section \ref{sec:summary} with a quick recap of the main result in \cite{Downing:2021mfw} and in section \ref{sec:phys} we explain its interpretation in terms of a system with a defect. 
This leads to an interpretation of the modular transform formulae in terms of an altered quantisation condition on the fermion modes which we explain in section \ref{sec:fqc}.
In section \ref{sec:manyI} we then generalise the results in \cite{Downing:2021mfw} for a single charge of spin 3 to an arbitrary \MD{but finite} combination of charges. \MD{In section \ref{sec:infinite} we extend the conjecture to the case with an infinite number of charges} and in section \ref{sec:R-} to the ``(R,-)'' sector which generalises the modular properties of the eta function.
Finally, in section \ref{sec:ham}, we present a construction of  the defect in a Hamiltonian formalism based on the methods of \cite{Toth:2006tj}, and then conclude with some conjectures and observations.


\section{Summary of previous results}
\label{sec:summary}

Our previous paper \cite{Downing:2021mfw} was concerned with the partition function for a free massless chiral fermion in the presence of a generalised Gibbs ensemble (GGE). The GGE was composed of two commuting quantities which can be thought of as the first two conserved charges of the KdV hierarchy in this model. If the fermion has modes $\psi_k$ then the KdV charges are
\begin{align}
    I_{2n-1}&= \sum_{k>0} k^{2n-1}\psi_{-k}\psi_k - c_{2n-1}^{\text{NS/R}}
    \;,
\end{align}
where $n$ is a positive integer, the sum is over $k={1,2,..}$ in the Ramond (R) sector and over $k=1/2,3/2..$ in the Neveu-Schwarz (NS) sector. The constant terms $c_{2n-1}^{\text{NS/R}}$ take values
\begin{align}
    c^{\NS/\R}_{2n-1} 
    \,=\, -\frac{(-1)^n}{(2\pi)^{2n}}
     \int_0^\infty\frac{t^{2n-1}}{1\pm e^t}\, \rd t
     \,=\,
    \begin{cases}
        \frac 12 (2^{1-2n}-1) \zeta(1-2n) & \mathrm{NS} \;,\\
        \frac 12 \zeta(1-2n) & \mathrm{R} \;,
    \end{cases}
    \label{eq:cformula}
\end{align}
where $\zeta(s)$ is the Riemann zeta function and the NS sector is given by $+$ and R by $-$. The fundamental quantities we considered were traces over the NS and R Fock spaces \begin{align}
  \chi^{\NS/\R,\pm}(\tau,\alpha) = 
  \Tr_{\NS/\R,\pm}\left( \;(\pm 1)^F\; q^{I_1} z^{I_3}\; \right)
  =
  \Tr_{\NS/\R,\pm}\left( \;(\pm 1)^F\; e^{2\pi i\,I_1} e^{\alpha\, I_3}\; \right)
  \;,
  \label{eq:GGETr}
\end{align}
where $F$ is the fermion number operator that has eigenvalue 1 on the ground state and anticommutes with $\psi_k$, and $z=e^\alpha$ is some complex parameter. 

{These traces can be evaluated explicitly to obtain $\chi^{R,-}(\tau,\alpha)=0$ and the others given by
\begin{align}
    \chi^{\NS,\pm}(\tau,\a)
    &=
    q^{-1/48}z^{7/1920}
    \prod_{n=1/2}
    \left(1 \pm q^n z^{k^3} \right)
    \;,\\
    \chi^{\R,+}(\tau,\a)
    &=
    2\,
    q^{1/24}z^{-1/240}
    \prod_{n=1}
    \left(1 + q^n z^{k^3} \right)
    \;.
\end{align}
}
Setting $\alpha=0$ in \eqref{eq:GGETr} we recover the standard expressions for free fermion characters\footnote{Note that $\chi^{R,+}(q) = 2 q^{1/24}+\ldots$ has a factor of two for the doubly degenerate ground state.}. 
If we consider a rectangular torus given by identifying the ends of a cylinder of circumference $L$ and length $R$, then the usual partition function is 
\begin{align}
    Z(L,R) 
    &
    = \frac 12 | \chi^{\NS,+}(\tau,0) |^2
    + \frac 12 | \chi^{\NS,-}(\tau,0) |^2
    + \frac 14 | \chi^{\R,+}(\tau,0) |^2
\;,    \\\;\;\;\;
q&=\exp(2\pi i \tau)\;,\;\;\tau = \frac {iR}L\;.\nonumber
\end{align}
Since this should not depend on the parametrisation of the torus, we have $Z(L,R)=Z(R,L)$. The individual functions $\chi^{\text{NS/R},\pm}$ are not invariant but satisfy
\begin{align}
\begin{pmatrix}
    \chi^{\NS,+}(\htau,0) \\ \chi^{\NS,-}(\htau,0) \\ \chi^{\R,+}(\htau,0)
\end{pmatrix}
=
\begin{pmatrix}
    1 & 0 & 0 \\ 0 & 0 & \frac 1{\sqrt 2} \\ 0 & {\sqrt 2} & 0
\end{pmatrix}
\begin{pmatrix}
    \chi^{\NS,+}(\tau,0) \\ \chi^{\NS,-}(\tau,0) \\ \chi^{\R,+}(\tau,0)
\end{pmatrix}
\;,\;\; \hq = \exp(2\pi i \htau)\;,\;\; \htau=-1/\tau\;.
\end{align}
This is a modular $S$ transformation: the maps $S:\tau\to-1/\tau$ and $T:\tau\to\tau+1$ generate the modular group $SL(2,Z)$. It was a natural question to ask if the functions \eqref{eq:GGETr} have similarly nice modular properties. 
The result we found in \cite{Downing:2021mfw} is ``no'', but instead
\begin{align}
\begin{pmatrix}
    \chi^{\NS,+}(\htau,\alpha) \\ \chi^{\NS,-}(\htau,\alpha) \\ \chi^{\R,+}(\htau,\alpha)
\end{pmatrix}
=
\begin{pmatrix}
    1 & 0 & 0 \\ 0 & 0 & \frac 1{\sqrt 2} \\ 0 & {\sqrt 2} & 0
\end{pmatrix}
\begin{pmatrix}
    \hat\chi^{\NS,+}(\tau,\alpha) \\ \hat\chi^{\NS,-}(\tau,\alpha) \\ \hat\chi^{\R,+}(\tau,\alpha)
\end{pmatrix}
\;,\;\; \hq = \exp(2\pi i \htau)\;,\;\; \htau=-1/\tau\;.
\label{eq:res3}
\end{align}
where
\begin{align}
\hat\chi^{\NS,+}(\tau,\alpha)
&=
q^{h_0^{\NS}(\tau,\alpha)}
\prod_{k=1/2}
\left (1 + e^{\tau x_1(k)}  \right)
\left (1 + e^{\tau x_2(k)}  \right)
\left (1 + e^{-\tau x_3(k)}  \right)\;,
\label{eq:NS+a2}
\\
\hat\chi^{\NS,-}(\tau,\alpha)
&= 
q^{h_0^{\NS}(\tau,\alpha)}
  \prod_{k=1/2}
\left (1 - e^{\tau x_1(k)}  \right)
\left (1 - e^{\tau x_2(k)}  \right)
\left (1 - e^{-\tau x_3(k)}  \right)\;,
\label{eq:R+a2}
\\
\hat\chi^{\R,+}(\tau,\alpha)
&
=
2\,
q^{h_0^{\R}(\tau,\alpha)}\,
 \left (1 + e^{\tau x_2(0)}  \right)\, \prod_{k=1}
\left (1 + e^{\tau x_1(k)}  \right)
\left (1 + e^{\tau x_2(k)}  \right)
\left (1 + e^{-\tau x_3(k)}  \right)\;.
\label{eq:NS-a2}
\end{align}
The ground state eigenvalues $h_0^{\NS/\R}(\t,\a)$ are defined by
\begin{align}
h_0^{\NS}(\t,\a) &= -\frac{1}{4 \pi^2} \int_0^\infty \frac{t}{e^t+1}
  f\left(-\frac{\alpha\tau^3t^2}{4\pisq}\right) \rd t
\;,\quad
  h_0^{\R}(\t,\a) = \frac{1}{4 \pi^2}\int_0^\infty \frac{t}{e^t-1}
  f\left(-\frac{\alpha\tau^3t^2}{4\pisq}\right) \rd t
  \;,
  \label{eq:h0def}    
\end{align}
where
\begin{align}
f(z)&={}_2F_1\left(\frac{1}{3},\frac{2}{3};\frac{3}{2};\frac{27}{8\pi i}z\right)
\;,
\label{eq:3.2.16}
\end{align}
and the roots $x_i(n)$ satisfy
\begin{equation}
x-\frac{\alpha\tau_2^3}{8\pi^3}x^3
=2n\pi i
\;,
\label{eq:5.2.7}
\end{equation}
where $n\in\mathbb{Z}$ for the R sector 
and
$n\in\mathbb{Z}+\frac{1}{2}$ for the NS sector. 
The roots of equation
\eqref{eq:5.2.7} are
\def\ftwoone{{F\!}}
\begin{align}
   x_1(n)
&= 2n\pi i\,
   \ftwoone\left(\tfrac{1}{3},\tfrac{2}{3};\tfrac{3}{2};-n^2\gamma\right)
\label{eq:5.2.8}
\\
&=2n\pi i
  \left(
   \tfrac{3}{2}\ftwoone\left(\tfrac{1}{3},\tfrac{2}{3};\tfrac{1}{2};1{+}n^2\gamma \right)
  {-}
   \tfrac{\sqrt{3}}{2}\sqrt{1{+}n^2\gamma}
   \ftwoone\left(\tfrac{5}{6},\tfrac{7}{6};\tfrac{3}{2};1{+}n^2\gamma\right)
  \right)
\;,
\nonumber
\\
   x_2(n)
&= - 2n\pi i
    \left(
    \tfrac{1}{2}\ftwoone\left(\tfrac{1}{3},\tfrac{2}{3};\tfrac{3}{2};-n^2\gamma\
\right)
   -\tfrac{1}{2n}\sqrt{-\tfrac{27}{\gamma}}
    \ftwoone\left(-\tfrac{1}{6},\tfrac{1}{6};\tfrac{1}{2};-n^2\gamma\right)
    \right)
\label{eq:5.2.10}
\\
&= -6n\pi i\,\ftwoone\left(\tfrac{1}{3},\tfrac{2}{3};\tfrac{1}{2};1+n^2\gamma\right)
\;,
\nonumber
\\
   x_3(n)
&= -2n\pi i
    \left(
   \tfrac{1}{2}\ftwoone
   \left(\tfrac{1}{3},\tfrac{2}{3};\tfrac{3}{2};-n^2\gamma\right)
   +\tfrac{1}{2n}\sqrt{-\tfrac{27}{\gamma}}\,
   \ftwoone\left(-\tfrac{1}{6},\tfrac{1}{6};\tfrac{1}{2};-n^2\gamma\right)
   \right)
\label{eq:5.2.12}
\\
&=2n\pi i\left(\tfrac{3}{2}\ftwoone\left(\tfrac{1}{3},\tfrac{2}{3};\tfrac{1}{2}\
;1{+}n^2\gamma\right)
{+}\tfrac{\sqrt{3}}{2}\sqrt{1{+}n^2\gamma}\,\ftwoone\left(\tfrac{5}{6},\tfrac{7\
}{6};\tfrac{3}{2};1{+}n^2\gamma\right)\right)
\;,
\nonumber
\end{align}
where
\be
 \ftwoone\,(a,b;c;z) \equiv{}_2F_1(a,b;c;z)\;,\;\; \hbox{ and }\;\;
 \gamma = \frac{27 \alpha \tau_2^3}{8\pi}
\;.
\ee
%
Note that $-x_3(k) = x_1(k)+x_2(k) = x_2(-k)$, so there are three
different ways to write the expressions (\ref{eq:NS+a2}-\ref{eq:NS-a2}).
Also, since $x_1(0)=0$ we have $(1+x^{\tau x_1(0)})=2$ and also $x_2(0)=-x_3(0)\
$, which is why we only include this term once in \eqref{eq:NS-a2}.

This result, \eqref{eq:res3}, which was conjectured in \cite{Downing:2021mfw} has now been proven by one of us [MD] in the companion paper \cite{downing2023modular}.
%
%
Having presented the result, we now turn to a physical interpretation.

\section{Physical interpretation in terms of defects}
\label{sec:phys}

In our original paper we did not have a satisfactory physical understanding of the modular transformed formulae. We now have one proposal which is both simple and elegant, which is to consider the original GGE not as a change in the action or Hamiltonian of the free fermion, but instead as due to the insertion of a line defect. We now present evidence for this interpretation.

\begin{figure}[htb]
\[
\begin{array}{lcl}
\multicolumn{3}{l}{
    \includegraphics[angle=0,width=10truecm]{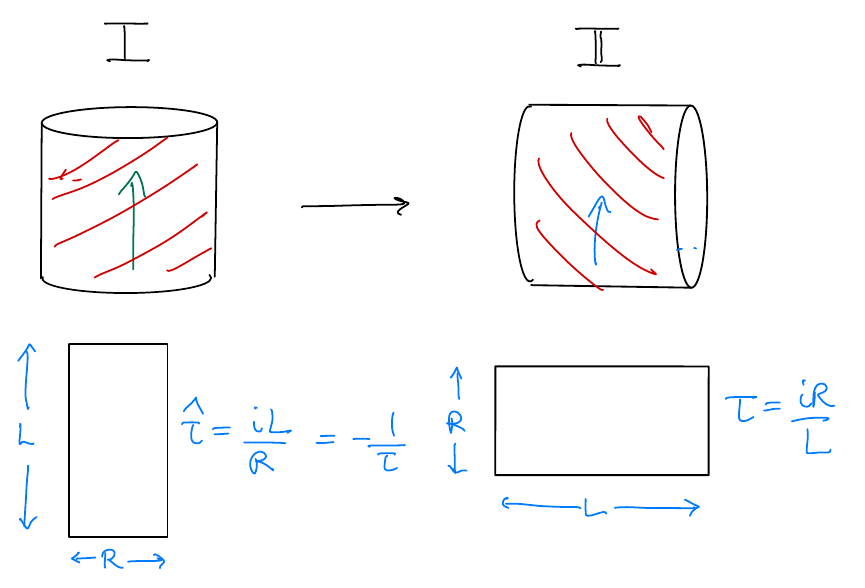}
    }
    \\[3mm]
    Z = \Tr(e^{-L H}) &\hspace{2truecm}&
    Z = \Tr(e^{-R H'})\\[1mm]
    H = \frac{2\pi}R (L_0 - c/24) - \frac{\alpha}L I_3 && H' =\;\; ?
\end{array}
\]    
\caption{Original interpretation of the modular transformed GGE traces.
\GW{We call (I) the direct channel and (II) the crossed channel.}}
    \label{fig:1}
\end{figure}

The original picture as envisaged in \cite{Downing:2021mfw} was that the GGE could be considered as the partition function for some system on a torus, expressed as a trace, as in figure \ref{fig:1}. 
The Hamiltonian $H'$ was unknown, as was the space on which it acts. All that is known are the ground state eigenvalues. From equations \eqref{eq:NS+a2}, \eqref{eq:R+a2} and \eqref{eq:NS-a2}, the ground state eigenvalues of $H'$ in the various sectors must be
\begin{align}
    E_0^{\NS/\R} &= \frac{2\pi}L\, h_0^{\NS/\R}(\t,\a)
    \nonumber\\&= 
    -\frac{1}{2\pi L}\int_0^\infty \frac{t}{1\pm e^t}f(-\alpha\tau^3t^2/4 \pisq)\,\rd t 
    \nonumber\\
    &= -\frac1{2\pi L}\int_0^\infty \log\left(1 \pm \exp\left(-u + \frac{\alpha R^3 u^3}{8 \picu L^3}\right)\right)\,\rd u
    \;,\label{eq:tbagse}
\end{align}
where again R is given by $-$ and NS by $+$. The last change of variables $t=u + \alpha\tau^3u^3/8 \picu i$ uses $\tau = iR/L$ and the fact that $f(z)$ satisfies 
\begin{align}
    f - \frac{z}{2\pi i}f^3 = 1\;.
\end{align}
This last integral \eqref{eq:tbagse} has the standard TBA form 
\begin{equation}
    LE_0^{\NS/\R} 
    = -\int_0^\infty \log(1 \pm e^{-\epsilon(u)}) \,\frac{\rd u}{2\pi}
\end{equation}
for a system of non-interacting massless particles with momentum $p=u/L$ and pseudo-energy 
\begin{equation}
    \epsilon(u) = u - \alpha u^3\left(\frac{R^3}{8\picu L^3}\right)
    \;,
\end{equation}
corresponding to the dispersion relation
\begin{align}
    E = p - \frac{\alpha}L \frac{p^3R^3}{8\picu}\;.
\end{align}
The minus sign in the R sector comes from introducing a ``twist" (see, for example \cite{Fendley_1992}), or, equivalently, a defect line with a transmission factor of $-1$.

This interpretation of the system on the torus (I) in figure \ref{fig:1}, \GW{(which we call the direct channel)} doesn't give any insight into the system on the torus (II) \GW{(which we call the crossed channel)} except that it should be some system with the opposite dispersion relation, swapping the roles of $E$ and $p$ \GW{(often called the mirror theory, \cite{Negro}).}

\begin{figure}[htb]
\[
\begin{array}{lcl}
\multicolumn{3}{l}{
    \includegraphics[angle=0,width=10truecm]{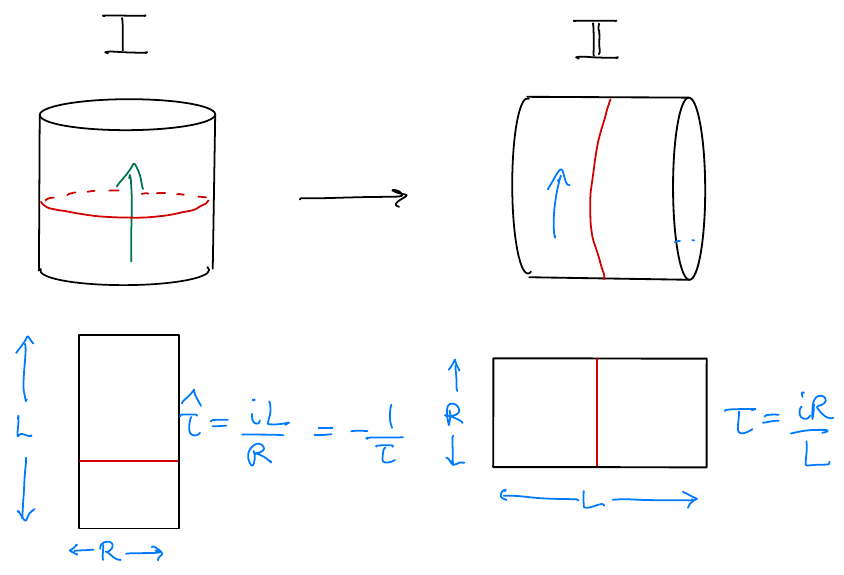}
    }
    \\[3mm]
    Z = \Tr(\;\hat D\;e^{-L H}) &\hspace{2truecm}&
    Z = \Tr(e^{-R H'})\\[1mm]
    H = \frac{2\pi}R (L_0 - c/24) && H' =\frac{2\pi}L(L_0 - c/24) + D(0)
\end{array}
\]    
\caption{Revised interpretation of the modular transformed GGE traces: On torus (I), the defect is inserted as an operator $\hat D$ in the trace; on torus (II) it is given by the addition of an operator $D(0)$ to the Hamiltonian.}
    \label{fig:2}
\end{figure}

There is, however, an alternative way to understand the ground state energies as TBA expressions, which is to read them as 
\begin{equation}
    LE_0 = -\int_0^\infty \log(1 + T(iu)\,e^{-\epsilon(u)}) \,\frac{\rd u}{2\pi}\;.
\label{eq:tba2}
\end{equation}
This is a system of non-interacting massless particles with the free massless pseudo-energy and dispersion relations
\begin{equation}
    \epsilon(u) = u\;,\;\; E = p = u/L\;,
\end{equation}
in the presence of a defect with transmission factor $T(u)$, as in figure \ref{fig:2}.

The TBA equations for a theory of a single massive particle with a defect insertion were first derived in \cite{Bajnok_2008}. It is very easy to derive the massless limit of these equations to get \eqref{eq:tba2}, which we do in appendix \ref{app:A}.

In the original calculation (the direct channel, torus (I)), this defect is placed along a line of constant time and corresponds to the insertion of an operator $\hat D$ in the trace; in the opposite (crossed) channel (torus (II)) it is placed along a line of constant position and its effect is instead to change the quantisation condition on the fermion momenta while leaving the dispersion relation unchanged. Formally we can consider the Hamiltonian in the crossed channel as 
\begin{align}
    H' = \frac{2\pi}L(L_0 - c/24) + D(0)
    \;,
    \label{eq:D0def}
\end{align}
where $D(0)$ is a local operator, or field, inserted at the location of the defect such that the spectrum of $H'$ agrees with the fermion quantisation condition.
This is shown in figure \ref{fig:2}.

\section{Fermion quantisation condtion}
\label{sec:fqc}

Comparing the ground state energy \eqref{eq:tbagse} found in \cite{Downing:2021mfw} with the TBA expression in the presence of a defect \eqref{eq:tba2} we see that the defect has transmission factor (with $p=u/L)$
\begin{align}
    T(u) 
    = \pm\exp({i\alpha u^3R^ 3/8\picu L^3})
    = \pm\exp({i\alpha R^3p^3/8\picu})
    \;.
    \label{eq:t(k)}
\end{align}
We consider the unperturbed system (without the defect) to be the fermion in the NS sector for which the momentum $p_n=(2\pi/L)k_n$ of the modes $\psi_{-k_n}$ satisfy
\begin{align}
    e^{ip_nL} = e^{2\pi i k_n} = -1
    \;,\;\;
    k_n = \tfrac 12, \tfrac 32, \ldots
    \;.
\end{align}
When the defect with transmission factor $T(u)$ is introduced into the system, the quantisation condition is altered to 
\begin{align}
    e^{ipL} T(u) = \pm e^{ipL + i\alpha R^3p^3/8\picu} = -1
    \label{eq:dqc}
\end{align}
so that
\begin{align}
   ip_nL + i\alpha(Rp_n/2\pi)^3 = 2n\pi i \;,\;\;\;\;
    \begin{cases}
        n\in \mathbb{Z} + \tfrac 12 & \NS\;, \\
        n\in \mathbb{Z}  & \R \;.
    \end{cases}
\end{align}
In terms of $x=ipL$ this is precisely the condition \eqref{eq:5.2.7}
\begin{align}
   x_n - \alpha\left(\frac{Rx_n}{2\pi L}\right)^3 = 2n\pi i \;,\;\;\;\;
    \begin{cases}
        n\in \mathbb{Z} + \tfrac 12 & \NS\;, \\
        n\in \mathbb{Z}  & \R \;,
    \end{cases}
\end{align}
and the condition that the fermion is actually right-moving is that the real part of $p$ is positive, so that the imaginary part of $x$ is positive.

This means that in the presence of the defect, the fermion is quantised with modes $\psi_{-k_n}$ for each solution of \eqref{eq:5.2.7} with positive imaginary part. Taking the standard expression for the partition function where each mode of energy $E_n$ contributes a factor
\begin{align}
(1 + e^{-RE_n}) = (1 + e^{-Rp_n}) = (1 + e^{iRx_n/L})= (1 + e^{\tau x_n})
\end{align}
to the partition function $Z=\Tr(e^{-RH})$, we get exactly the results \eqref{eq:NS+a2}, \eqref{eq:R+a2} and \eqref{eq:NS-a2}.



%

These formulae do not represent the trace over the product of three separate sets of fermion modes, as we conjecture in \cite{Downing:2021mfw}, but instead the trace over the set of modes of a single fermion subject to the deformed quantisation condition \eqref{eq:dqc}.

\section{Extending the result to more charges}
\label{sec:manyI}
In \cite{Downing:2021mfw} we gave a conjecture for the exact transformation of the GGE with only the $I_3$ KdV charge inserted. Here we first extend the transformation results to the case of a single charge $I_{2m-1}$ and then with a arbitrary but finite collection of $I_{2m-1}$ in the GGE. These conjectures have been proved in the companion paper \cite{downing2023modular}. Finally in section \ref{sec:infinite}, we extend the conjecture to the case with an infinite number of charges inserted. Here there are issues of convergence for the GGE so the result is more speculative but it is a natural generalisation from the finite case.


The starting point of the derivation of the formulae 
\eqref{eq:res3}-\eqref{eq:NS-a2} in the case of a GGE with the single charge 
$I_3$ was the expression \eqref{eq:h0def} for the ground state eigenvalue of $H'$ in the crossed channel. This formula was found by summing over the ground state energies in a putative expression for the transformed GGE - 
\begin{align}
    \Tr_{\NS,\R}(\hq^{L_0 - c/24} e^{\alpha I_3})
    \sim \Tr(q^{L_0 - c/24} e^{\sum_{n=1}^\infty \alpha'_{2n+1} I_{2n+1}})
    = q^{h-c/24}\, e^{\sum_{n=1}^\infty \alpha'_{2n+1} c^{\NS/\R}_{2n+1}}(1 + \ldots)\;.
    \label{eq:asymp}
\end{align}
While \eqref{eq:asymp} is only asymptotically true, it is possible to find the $\alpha_{2n+1}'$ relatively easily and, using the integral representation \eqref{eq:cformula} for $c_{2n-1}$, find the expressions
\begin{align}
    h_0^{\NS/\R}
    &=h -c/24 + \sum_{n=1}^\infty\frac{\alpha'_{2n+1}}{2\pi i \tau}\,c_{2n+1}^{\NS/\R}
    \nonumber\\
    &=c_1^{\NS/\R} + \sum_{n=1}^\infty\frac{\alpha'_{2n+1}}{2\pi i \tau}\,c_{2n+1}^{\NS/\R}
    \nonumber\\
    &=
 -\frac{1}{4\pi^2}\int_0^\infty du \log(1  \pm e^{-u -\a(iu\tau/2\pi)^{3}})
 \;,
\label{eq:e0}
\end{align}
leading to the excited state energies being found either from analytic continuation of \eqref{eq:e0}, or from the quantisation condition in the presence of the defect with transmission factor $T(iu)= \pm\exp(-\alpha(iu\tau/2\pi)^3)$

It is straightforward to use the results of \cite{Downing:2021mfw} to repeat these calculations for a GGE with a single higher charge $I_{2m-1}$. The corresponding ground state energies $h_0^{\NS/\R}$ can be found from equations (6.15) and (6.16) in \cite{Downing:2021mfw}, as  
\begin{align}
    h_0^{\NS/\R}
    &=h -c/24 + \sum_{n=1}\frac{\alpha'_{2n+1}}{2\pi i \tau}\,c_{2n+1}^{\NS/\R}
    \nonumber\\
    &=
 -\frac{1}{4\pi^2}\int_0^\infty dx \log(1  \pm e^{-x -\a(ix\tau/2\pi)^{2m-1}})
 \;,
\label{eq:e02}
\end{align}
which again can be interpreted as due to a defect with transmission factor 
\begin{align}
    T(iu)= \pm\exp(-\alpha(iu\tau/2\pi)^{2m-1})
    \;.
\end{align}
We can now make an obvious conjecture for the general case:
if the GGE is composed of a finite set of charges,
\begin{align}
    \alpha_3 I_3 + \ldots \alpha_{2m+1}I_{2m+1}
    = \sum_{p=1}^m \alpha_{2p+1} I_{2p+1}
\end{align}
then it is equivalent to a defect with transmission factor
\begin{align}
    T(iu) = \pm \exp\left( - \sum_{p=1}^m \alpha_{2p+1}(iu\tau/2\pi)^{2p+1} \right)\;.
\end{align}

We can now repeat all the previous arguments on fermion mode quantisation and end up with the following transformation formulae.
We denote the set of chemical potentials $\{\alpha_3,\ldots,\alpha_{2m+1}\}$ by $\bm\alpha$ and define the polynomial
\begin{align}
    p^{(\ba)}(x) = \sum_{p=1}^m \alpha_{2p+1} x^{2p+1}
    \;.
\end{align}
The GGE traces are then
\begin{align}
    \chi^{\NS,\pm}(\tau,\ba)
    &=
    q^{-1/48}e^{-\sum_{p=1}^m \a_{2p+1}c_{2p-1}^\text{NS}}
    \prod_{n=1/2}
    \left(1 \pm q^n e^{ p^{(\ba)} (n)} \right)
    \;,\\
    \chi^{\R,+}(\tau,\ba)
    &=
    2\,
    q^{1/24}e^{-\sum_{p=1}^m \a_{2p+1}c_{2p-1}^\text{R}}
    \prod_{n=1}
    \left(1 + q^n e^{ p^{(\ba)} (n)} \right)
    \;,
\end{align}
and their modular transforms are
\begin{align}
\begin{pmatrix}
    \chi^{\NS,+}(\htau,\ba) \\ \chi^{\NS,-}(\htau,\ba) \\ \chi^{\R,+}(\htau,\ba)
\end{pmatrix}
=
\begin{pmatrix}
    1 & 0 & 0 \\ 0 & 0 & \frac 1{\sqrt 2} \\ 0 & {\sqrt 2} & 0
\end{pmatrix}
\begin{pmatrix}
    \hat\chi^{\NS,+}(\tau,\ba) \\ \hat\chi^{\NS,-}(\tau,\ba) \\ \hat\chi^{\R,+}(\tau,\ba)
\end{pmatrix}
\;,\;\; \hq = \exp(2\pi i \htau)\;,\;\; \htau=-1/\tau\;.
\label{eq:res3ba}
\end{align}
where
\begin{align}
\hat\chi^{\NS,\pm}(\tau,\ba)
&=
q^{h_0^{\NS}(\tau,\ba)}
\prod_{k\in\Z+\frac 12}
\prod_{\substack{x_j(k)\\\text{Im}(x_j(k))>0}} 
\left(1 \pm e^{\tau x_j(k)}\right)\;,
\label{eq:NS+a2ba}
\\
\hat\chi^{\R,+}(\tau,\ba)
&
=
2q^{h_0^{\R}(\tau,\ba)}\,
\prod_{k\in\Z}
\prod_{\substack{x_j(k)\\\text{Im}(x_j(k))>0}} 
\left(1 + e^{\tau x_j(k)}\right)
\;.                        
\label{eq:NS-a2ba}
\end{align}
The ground state eigenvalues $h_0^{\NS/\R}(\t,\a)$ are given by
\begin{align}
h_0^{\NS/\R}(\t,\a) &= -\frac{1}{4 \pi^2} \int_0^\infty 
\rd u
\log(1 \pm e^{-u-p^{(\ba)}(iu\tau/2\pi)})
  \;,
  \label{eq:h0defba}    
\end{align}
and the roots $x_i(k)$ satisfy
\begin{equation}
x + p^{(\ba)}(i \tau x/2\pi)
=2n\pi i
\;,
\label{eq:polynomialba}
\end{equation}
where $n\in\mathbb{Z}$ for the R sector 
and
$n\in\mathbb{Z}+\frac{1}{2}$ for the NS sector. 
This transformation formula has been proven in \cite{downing2023modular}.

Equation \eqref{eq:polynomialba} is the quantisation condition for the one particle energies in the presence of the line defect.
In section \ref{sec:ham} we reproduce the quantisation condition \eqref{eq:polynomialba} by constructing the defect Hamiltonian explicitly.

\section{Infinite collection of charges}\label{sec:infinite}

We can extend the conjectured transform even further to the case where we have an infinite number of charges inserted. This case was also commented on in \cite{downing2023modular} but the transformation was not proved there. We will discuss the issues involved with proving these transforms at the end of this section.

Formally the GGE with all charges inserted is
\be\label{eq:FullGGE}
    \Tr_{\text{NS/R},\pm}\left(e^{ \sum_{n=2}^\infty \a_{2n-1}I_{2n-1}} q^{I_1}\right) = e^{-\sum_{n=2}^\infty\alpha_{2n-1}c_{2n-1}^{\text{NS/R}}}q^{-c_1^{\NS/\R}}\prod_{k\geq 0}\left(1 \pm e^{\sum_{n=2}^\infty \a_{2n-1}k^{2n-1}} q^k\right)\;.
\ee
Since this contains both infinite series and products there are potential convergence issues. We must restrict the domain of the chemical potentials, $\a_{2n-1}$, so that the right hand side of \eqref{eq:FullGGE} is well defined. We have to consider the convergence of the two series
\be
    2\pi i\t k + \sum_{n=2}^\infty \a_{2n-1}k^{2n-1} \quad\text{and}\quad 2\pi i \t c_1^{\NS/\R} + \sum_{n=2}^\infty\alpha_{2n-1}c_{2n-1}^{\text{NS/R}}\;,
\ee
and the convergence of the product. Since the product in \eqref{eq:FullGGE} is over all \MD{$k \in \Z$ or $\Z+\frac{1}{2}$ with $k\geq0$}, the radius of convergence of the series $2\pi i\t k + \sum_{n=1}^\infty \a_{2n-1}k^{2n-1}$ must be infinite. Additionally, as $k$ increases, the value of the exponential of the series must decay sufficiently fast for the product to converge. We also need the series $2\pi i\t c_1^{\NS/\R} + \sum_{n=1}^\infty\alpha_{2n-1}c_{2n-1}^{\text{NS/R}}$ to converge. These conditions together constrain the allowed values of the chemical potentials if we want \eqref{eq:FullGGE} to be well defined.

Alternatively if the right hand side is not well defined for our choice of $\a_{2n-1}$ we can obtain a well defined expression through a regularisation process. We first define the function $f_{\t,\ba}(k)$ as the power series expansion
\be
    f_{\t,\ba}(k) = 2\pi i\t k + \sum_{n=2}^\infty \a_{2n-1}k^{2n-1} \;,\quad |k|<R\;,
\ee
which has radius of convergence $R$. If $R$ is finite then as we take the product over $k$ in \eqref{eq:FullGGE} there is a point where $k>R$ so the power series will no longer converge. However if the function $f_{\t,\ba}(k)$ is defined for all $k>0$ when we can replace the power series with $f$ in the product. We still need to check that the product converges. 

There is also the series $2\pi i\t c_1^{\NS/\R} + \sum_{n=2}^\infty \a_{2n-1}c^{\text{NS/R}}_{2n-1}$ to consider. If this series does not converge for our choice of chemical potentials we can again regularise this series as follows. Using the integral representation \eqref{eq:cformula} for $c^{\text{NS/R}}_{2n-1}$ the series can be written as
\be\begin{split}
    &\sum_{n=1}^\infty \a_{2n-1}c^{\text{R}}_{2n-1} = \int_0^\infty \frac{dt}{2\pi i} \frac{f_{\t,\ba}\left(\frac{t}{2\pi i}\right)}{e^t-1}\;,\\
    &\sum_{n=1}^\infty \a_{2n-1}c^{\text{NS}}_{2n-1} = -\int_0^\infty \frac{dt}{2\pi i} \frac{f_{\t,\ba}\left(\frac{t}{2\pi i}\right)}{e^t+1}\;.
\end{split}\ee
Again this is a formal relation since if the power series of $f_{\t,\ba}(t)$ has a finite radius of convergence and we are integrating over $t>0$ there will be a point where the expansion is no longer valid and the integral and sum cannot be swapped. We need the integrals to converge for the GGE to be defined but this is a weaker condition then the convergence of the original sum. 

Using these regularisations we will take the GGEs to be defined as
\be\begin{split}
    &\chi^{\R,+}(\tau,\ba) = \Tr_{\text{R},+}\left(e^{\sum_{n=2}^\infty \a_{2n-1}I_{2n-1}} q^{I_1}\right) = \exp\left(-\int_0^\infty \frac{dt}{2\pi i} \frac{f_{\t,\ba}\left(\frac{t}{2\pi i}\right)}{e^t-1}\right) \prod_{k = 0}\left(1 + e^{f_{\t,\ba}(k)}\right)\;,\\
    &\chi^{\NS,\pm}(\tau,\ba) = \Tr_{\text{NS},\pm}\left(e^{ \sum_{n=2}^\infty \a_{2n-1}I_{2n-1}} q^{I_1}\right) = \exp\left(\int_0^\infty \frac{dt}{2\pi i} \frac{f_{\t,\ba}\left(\frac{t}{2\pi i}\right)}{e^t+1}\right) \prod_{k = \frac{1}{2}}\left(1 \pm e^{f_{\t,\ba}(k)}\right)\;.
\end{split}\ee
We can now extend the transformation for the finite case \eqref{eq:res3ba} to a conjectured transform for the infinite case
\begin{align}
\begin{pmatrix}
    \chi^{\NS,+}(\htau,\ba) \\ \chi^{\NS,-}(\htau,\ba) \\ \chi^{\R,+}(\htau,\ba)
\end{pmatrix}
=
\begin{pmatrix}
    1 & 0 & 0 \\ 0 & 0 & \frac 1{\sqrt 2} \\ 0 & {\sqrt 2} & 0
\end{pmatrix}
\begin{pmatrix}
    \hat\chi^{\NS,+}(\tau,\ba) \\ \hat\chi^{\NS,-}(\tau,\ba) \\ \hat\chi^{\R,+}(\tau,\ba)
\end{pmatrix}
\;,\;\; \hq = \exp(2\pi i \htau)\;,\;\; \htau=-1/\tau\;,
\end{align}
where
\begin{align}
\hat\chi^{\NS,\pm}(\tau,\ba)
&=
q^{h_0^{\NS}(\tau,\ba)}
\prod_{k\in\Z+\frac 12}
\prod_{\substack{x_j(k)\\\text{Im}(x_j(k))>0}} 
\left(1 \pm e^{\tau x_j(k)}\right)\;,
\\
\hat\chi^{\R,+}(\tau,\ba)
&
=2^N q^{h_0^{\R}(\tau,\ba)}\,
\prod_{k\in\Z}
\prod_{\substack{x_j(k)\\\text{Im}(x_j(k))>0}} 
\left(1 + e^{\tau x_j(k)}\right)
\;.
\end{align}
The ground state eigenvalues $h_0^{\NS/\R}(\t,\a)$ are given by
\begin{align}
h_0^{\NS/\R}(\t,\a) &= -\frac{1}{4 \pi^2} \int_0^\infty 
\rd u
\log(1 \pm e^{-f_{\hat\t,\ba}(iu\tau/2\pi)})
  \;,   
\end{align}
the functions $x_i(k)$ satisfy
\begin{equation}
f_{\hat\t,\ba}\left(\frac{i\t x}{2\pi}\right)
=2\pi i k
\;,
\end{equation}
where $k\in\mathbb{Z}$ for the R sector and $k\in\mathbb{Z}+\frac{1}{2}$ for the NS sector and $N$ is the number of $x_i(k)$ that vanish at $k=0$.

Since we do not have an explicit expression for the function $f_{\t,\ba}(x)$ we cannot say much about the solutions $x_i(k)$. When proving the transform in \cite{downing2023modular} for the finite case the $x_i(k)$ are roots of a polynomial. This allowed us to derive several properties of the roots that were then used in the proof. It would be interesting to consider the cases where we have a set of the chemical potentials were infinitely many of the $\a_{2n-1}$ are non-zero so that we have a known function $f_{\t,\ba}(x)$. If the $x_i(k)$ are also known functions then we can potentially prove the transform in these special cases.

\section{The ``$(R,-)$'' sector}
\label{sec:R-}

In this section we will also consider the sector ``(R,$-$)'' in which the fermion field is periodic on both cycles of the torus and so its contribution is calculated by taking a trace in the Ramond sector with the insertion of $(-1)^F$. Since the periodicity is the same on both cycles, \GW{the sector}
is invariant under $\tau\to-1/\tau$ which interchanges the cycles.

While this sector does contribute to correlation functions in the conformally invariant Ising model, for example the one-point function of the energy operator, it gives zero contribution to the partition function in the Ising model.
The reason is that the the full Ising model is constructed from left- and right- moving fermions. The relevant state space is then in the $(R,R)$ sector, with both left- and right- moving fermions. The presence of two anticommuting fermion zero modes $\{\psi_0,\bar\psi_0\}=0$ means the ground state space is two-dimensional with equal numbers of bosonic and fermionic states. The same is true at every level in the $(R,R)$ sector and so the trace $\Tr_{(R,R)}(\,(-1)^F e^{-RH})$ contributes zero to the partition function.

The same equality of bosonic and fermion states also occurs when we consider just the left-moving sector but want to include both $(-1)^F$ and $\psi_0$. Since these also anticommute, $\{\psi_0,(-1)^F\}=0$, there must be at least a two-dimensional highest weight space with equal numbers of bosonic and fermionic states and so
\begin{align}
\Tr_{\R}\left(\,(-1)^F q^{L_0 - c/24}\right) =0 
\;.
\end{align}
We can, however, consider a restricted fermion algebra which does not include the fermion zero mode, and a corresponding state space with a one-dimensional highest weight space $R'$ spanned by a bosonic state of conformal weight $1/16$. In this case the traces are simply
\begin{align}
\Tr_{\R'}\left(\,q^{L_0-c/24}\,\right)
 &= q^{1/24} \prod_{n>0} (1 + q^n)
 =\tfrac12 \chi^{\R,+}(q)\;,
 \\
\Tr_{\R'}\left(\,(-1)^F\,q^{L_0-c/24}\,\right)
 &= q^{1/24} \prod_{n>0} (1 - q^n)= \eta(\tau)
 \;,
\end{align}
where $\eta(\tau)$ is the usual Dedekind eta function which transforms as a modular form of weight 1/2:
\begin{align}
    \eta(-1/\tau) = \sqrt{-i\tau}\,\eta(\tau)
    \;.
\end{align}
While this function is not invariant, as might have been hoped for from the previous discussion of periodicities -- a fact we can attribute to the omission of the fermion zero mode -- it certainly has well-defined modular properties.
It is then straightforward to extend all the considerations of the previous sections to traces in the $(R',-)$ sector, with the result (again proven in \cite{downing2023modular}):
\begin{align}
\chi^{\R,-}(\htau,\ba)
    &=\sqrt{-i\tau} \,\hat\chi^{\R,-}(\tau,\ba)
    \;,\;\;
\end{align}
where the functions $\chi$ and $\hat\chi$ are
\begin{align}
     \chi^{\R,-}(\tau,\ba)
    &=
    q^{1/24}
    \prod_{n=1}
    \left(1 - q^n e^{ p^{(\ba)} (n)} \right)
    \;,\\
    \hat\chi^{\R,-}(\tau,\ba)
&
=
q^{h_0^{\R}(\tau,\ba)}\,
\prod_{k\in\Z}
\prod_{\substack{x_j(k)\\\text{Im}(x_j(k))>0}} 
\left(1 - e^{\tau x_j(k)}\right)
\;,
\end{align}
and the roots $x_i(k)$ satisfy
\eqref{eq:polynomialba}.

\section{Defect Hamiltonian}
\label{sec:ham}

In section \ref{sec:phys}, we stated that formally we can consider the Hamiltonian in the crossed channel to be the perturbation of the usual conformal Hamiltonian by a local field $D(0)$ such that the spectrum of the Hamiltonian matches the quantisation condition on the fermion modes.
In this section we present one construction of just such a modified Hamiltonian.

If we consider, as before, just the insertion of $I_3$, then we meet the immediate problem that the line defect in the original (direct) channel can be considered as the integral of an irrelevant field,
\begin{align}
    e^{\alpha I_3}
    \;,\;\;
    I_3 = \left(\tfrac{L}{2\pi}\right)^3 \int_0^L 
    \tfrac 12 (\psi'''(x)\psi(x))\,\frac{\rd x}{2\pi}
    =
    \left(\tfrac{L}{2\pi}\right)^3 \int_0^L 
    \tfrac 67 (T(x)T(x))\,\frac{\rd x}{2\pi}
    \;.
\end{align}
Our first guess for \GW{the perturbing field} $D(x)$ will be that same irrelevant field. However, perturbations by irrelevant fields are well known to be plagued by divergences and so we have used an approach which seems to avoid such problems: this is the method of G.Zs.~T\'oth \cite{Toth:2006tj} which proved very effective for the boundary perturbation of the free fermion by the irrelevant field $T(x)$ without the need for any further explicit regularisation or renormalisation. The effect of the perturbation was to induce a boundary reflection factor which (as with the defect transmission factor here) modified the fermion quantisation condition. The construction of \cite{Toth:2006tj} added a boundary field to the Hamiltonian which led to altered equations of motion for the fermion which in turn led to altered boundary conditions and reproduced precisely the correct quantisation condition. In this section we attempt to repeat that idea for the defect transmission factor. 


Throughout this section we will be working with space, $x$, and time, $t$, coordinates rather than the complex coordinates $(z,\bar z)$ used previously.

We start with a free right moving fermion $\psi(x,t)$. The field $\psi$ is defined on the cylinder with $x\in [0,L]$ and $t\in\mathbb{R}$. We impose either periodic $(+)$ or anti periodic $(-)$ boundary conditions in the spatial direction
\be\label{eq:ffbndry}
    \psi(0,t)=\pm\psi(L,t)\;,
\ee
and the equal time anti-commutator between two fields is
\be\label{eq:ffanticom}
    \{\psi(x,t),\psi(y,t)\} = -2\pi i\,\delta(x-y)\;.
\ee
Recall that the Hamiltonian for the free fermion is
\be\label{eq:freehamiltonian}
    H_0 = -\frac{1}{4\pi}\int_0^{L}\dx\, \partial_x \psi\, \psi = -\int_0^L \frac{\dx}{2\pi} T(x) = (L_0 - c/24)\;.
\ee

We could introduce the line defect directly into the right-moving fermion on the line at $x=L/2$, but instead we use the folding trick to map it to an equivalent system of a left- and a right-moving fermion on a strip with a trivial boundary condition at $x=0$ and a non-trivial boundary condition at $L/2$. We do this since it enables us to use precisely the same method as \cite{Toth:2006tj}, as follows:

First, fold the line $\frac{L}{2} < x < L$ over onto the line $0<x<\frac{L}{2}$, so the point $x$ is identified with $L-x$. Now we define two new fermion fields $\Phi_i(x,t)$, $i=1,2$, that both live on the line $0<x<\frac{L}{2}$, in terms 
\GW{of} 
the original field $\psi(x,t)$
\be\label{eq:phipsi}
    \Phi_1(x,t) = \sqrt{\frac{iL}{\pi}}\psi(x,t) \;,\quad \Phi_2(x,t) = \sqrt{\frac{iL}{\pi}}\psi(L-x,t)\;.
\ee
We have chosen the normalisation of $\Phi_i$ to match the conventions in \cite{Toth:2006tj}. Note that in \cite{Toth:2006tj} the fields $\Phi_i(x,t)$ are defined on the cylinder $(x,t)\in[0,L]\times\mathbb{R}$ while our $\Phi_i(x,t)$ are defined on $(x,t)\in[0,L/2]\times\mathbb{R}$ so the commutators and Hamiltonian differ from those in \cite{Toth:2006tj} by factors of 2. Since $\psi$ is right-moving on the full line, after the folding $\Phi_1$ is right-moving and $\Phi_2$ is left-moving. From \eqref{eq:ffbndry} the new fields have the boundary conditions
\be\label{eq:Phibc}
    \Phi_1(0,t) = \pm \Phi_2(0,t) \;,\quad \Phi_1\left(\tfrac{L}{2},t\right) = \Phi_2\left(\tfrac{L}{2},t\right)\;.
\ee
Additionally the anti-commutator \eqref{eq:ffanticom} becomes
\be\label{eq:Phianticom}\begin{split}
    &\{\Phi_1(x,t),\Phi_1(y,t)\} = \{\Phi_2(x,t),\Phi_2(y,t)\} = 2L\delta(x-y)\;,\\
    &\{\Phi_1(x,t),\Phi_2(y,t)\} =  2L( \delta(x+y-L) \pm \delta(x+y) )\;.
\end{split}\ee
The second anti-commutation relation comes from the two boundaries where the fermion fields are identified as in \eqref{eq:Phibc}. The Hamiltonian for the free fermions is now
\be
    H_0 = -\frac{i}{4L}\int_0^{L/2} \dx\,(\Phi_1\partial_x\Phi_1 - \Phi_2\partial_x\Phi_2)\;.
\ee
We will use the Hamiltonian formalism to compute the equations of motion 
\be\label{eq:eom}
    \partial_t\Phi_i(x,t) = i[H_0,\Phi_i(x,t)]\;.
\ee
However before we do this we introduce our conventions for the Dirac delta function and the Heaviside step function. When integrated against a test function the Dirac delta function gives the following results for $a>0$
\be
    \int_{-a}^a \dx\, \delta(x)f(x)=f(0)\;,\quad \int_0^a \dx\, \delta(x)f(x) = \int_{-a}^0 \dx\, \delta(x)f(x) = \frac{1}{2}f(0)\;.
\ee
The Heaviside step function is defined as the integral of the Dirac delta function
\be
    \Theta(x)=\int_{-\infty}^x\dx'\,\delta(x')\;,
\ee
so it takes the values
\be
    \Theta(x)=\begin{cases}0,&x<0\;,\\\tfrac{1}{2},&x=0\;,\\1,&x>0\;.\end{cases}
\ee
Hence if we integrate $\delta(x-y)$ against a test function $f(x)$ in the interval $[a,b]$ then
\be
    \int_a^b \dx f(x)\delta(x-y) = f(y)(1 - \Theta(a-y) - \Theta(y-b))\;,
\ee
and if we integrate $\delta'(x-y)$ against a test function then
\be
    \int_a^b \dx f(x)\delta'(x-y) = \delta(y-b)f(b) - \delta(y-a)f(a) - f'(y)(1 - \Theta(a-y) - \Theta(y-b))\;.
\ee
We use these results and the anti-commutators \eqref{eq:Phianticom} to compute the equations of motion \eqref{eq:eom}
\be\begin{split}
    &\partial_t\Phi_1(x,t) = -\partial_x\Phi_1(x,t) - \frac{1}{2}\delta(x)(\Phi_1(0,t) \mp \Phi_2(0,t)) +\frac{1}{2} \delta\left(x - \tfrac{L}{2}\right) \left(\Phi_1\left(\tfrac{L}{2},t\right)-\Phi_2\left(\tfrac{L}{2},t\right)\right) \\
    & +\Theta(-x)(\partial_x\Phi_1(0,t) \pm \partial_x\Phi_2(0,t)) + \Theta\left(x - \tfrac{L}{2}\right) \left(\partial_x\Phi_1\left(\tfrac{L}{2},t\right)+\partial_x\Phi_2\left(\tfrac{L}{2},t\right)\right)\;,
\end{split}\ee
and
\be\begin{split}
    &\partial_t\Phi_2(x,t) = \partial_x\Phi_2(x,t) + \frac{1}{2}\delta(x)(\Phi_2(0,t) \mp \Phi_1(0,t)) +\frac{1}{2} \delta\left(x - \tfrac{L}{2}\right) \left(\Phi_1\left(\tfrac{L}{2},t\right)-\Phi_2\left(\tfrac{L}{2},t\right)\right) \\
    & -\Theta(-x)(\partial_x\Phi_2(0,t) \pm \partial_x\Phi_1(0,t)) - \Theta\left(x - \tfrac{L}{2}\right) \left(\partial_x\Phi_1\left(\tfrac{L}{2},t\right)+\partial_x\Phi_2\left(\tfrac{L}{2},t\right)\right)\;,
\end{split}\ee
The delta function terms $\delta(x)$ and $\delta\left(x - \frac{L}{2}\right)$ in the above equations vanish by the boundary conditions \eqref{eq:Phibc}. So we have
\be\begin{split}
    \partial_t\Phi_1(x,t) =& -\partial_x\Phi_1(x,t) +\Theta(-x)(\partial_x\Phi_1(0,t) \pm \partial_x\Phi_2(0,t)) \\
    &+ \Theta\left(x - \tfrac{L}{2}\right) \left(\partial_x\Phi_1\left(\tfrac{L}{2},t\right)+\partial_x\Phi_2\left(\tfrac{L}{2},t\right)\right)\;,\\
    \partial_t\Phi_2(x,t) =& \partial_x\Phi_2(x,t) -\Theta(-x)(\partial_x\Phi_2(0,t) \pm \partial_x\Phi_1(0,t)) \\
    &- \Theta\left(x - \tfrac{L}{2}\right) \left(\partial_x\Phi_1\left(\tfrac{L}{2},t\right)+\partial_x\Phi_2\left(\tfrac{L}{2},t\right)\right)\;,
\end{split}\ee
Solving these equations will give the usual free fermion solution and quantisation condition 
(as we will see below). 

Now we add a line defect at $x=\frac{L}{2}$. We do this by adding terms $H_D^{(2n-1)}$, $n=1,2,\dots$, to the Hamiltonian $H_0$ to get
the full Hamiltonian for the system with a line defect:
\be\label{eq:FullH}
    H=H_0+\sum_{n=1}\lambda_{2n-1}H_D^{(2n-1)}\;,
\ee
and the equations of motion for the full system are
\be
    \partial_t\Phi_i(x,t) = i[H,\Phi_i(x,t)]\;.
\ee
We take the additional terms $H_D^{(2n-1)}$ to be
\be\label{eq:defectops}
    H_D^{(2n-1)}=i\left( \Phi_1\left(\tfrac{L}{2},0\right) + \Phi_2\left(\tfrac{L}{2},0\right)\right) \lim_{y\rightarrow \frac{L}{2}}\partial_y^{2n-1}(\Phi_2(y,0)-\Phi_1(y,0))\;.
\ee
The limit regularises later expressions that would otherwise contain delta function singularities. It removes these singularities by using the prescription
\be\label{eq:deltaprescrip}
    \lim_{y\rightarrow \frac{L}{2}}\delta^{(n)}\left(y-\tfrac{L}{2}\right) = 0\;,\quad n=0,1,\dots\;.
\ee
Throughout the calculation we will keep the limit explicit and only take it at the end. The equations of motion of the full system contain the commutator of the defect terms $H_D^{(2n-1)}$ with the fields $\Phi_i$
\be
    [H_D^{(2n-1)},\Phi_i(x,t)]=-4iL\delta\left(x - \tfrac{L}{2}\right)\lim_{y\rightarrow \frac{L}{2}}\partial_y^{2n-1}(\Phi_2(y,t)-\Phi_1(y,t))\;,\quad i=1,2\;.
\ee
This means the equations of motion for $\Phi_i$
are now
\be\label{eq:Phi1eom}\begin{split}
    &\partial_t\Phi_1(x,t) + \partial_x\Phi_1(x,t) \\
    &= \Theta(-x)(\partial_x\Phi_1(0,t) \pm \partial_x\Phi_2(0,t)) + \Theta(x-L/2)(\partial_x\Phi_1(L/2,t)+\partial_x\Phi_2(L/2,t))\\
    &+4L\delta(x-L/2)\sum_{n=1}\lambda_{2n-1}\lim_{y\rightarrow L/2}\partial_y^{2n-1}(\Phi_2(y,t)-\Phi_1(y,t))\;,\\
\end{split}\ee
\be\label{eq:Phi2eom}\begin{split}
    &\partial_t\Phi_2(x,t) - \partial_x\Phi_2(x,t) \\
    &=  - \Theta(-x)(\partial_x\Phi_2(0,t) \pm \partial_x\Phi_1(0,t)) - \Theta(x-L/2)(\partial_x\Phi_1(L/2,t)+\partial_x\Phi_2(L/2,t))\\
    &+4L\delta(x-L/2)\sum_{n=1}\lambda_{2n-1}\lim_{y\rightarrow L/2}\partial_y^{2n-1}(\Phi_2(y,t)-\Phi_1(y,t))\;.
\end{split}\ee
We will use the equations of motion \eqref{eq:Phi1eom} and \eqref{eq:Phi2eom} to derive the quantisation condition for the single particle energies. Since the Hamiltonian is quadratic in the fields this is still a ``free'' fermion and hence the full spectrum is composed from the sum of multiple single particle states. Hence if we know the quantisation condition on the single particle states we know the full spectrum. 

To find the one particle energies start with the eigenstates $|E_1\rangle$ and $|E_2\rangle$ of the full Hamiltonian $H$ \eqref{eq:FullH}. We will derive the quantisation condition from the matrix elements
\be
    \langle E_1|\Phi_i(x,t)|E_2\rangle \;,\quad i=1,2\;.
\ee
In the Heisenberg picture, the time evolution of the fields is
\be
    \Phi_i(x,t)=e^{itH}\Phi_i(x,0)e^{-itH}\;.
\ee
Since the two states are eigenstates of $H$ we find
\be\label{eq:matrixel}
    \langle E_1|\Phi_i(x,t)|E_2\rangle=\langle E_1|\Phi_i(x,0)|E_2\rangle e^{ikt} = f_i(x)e^{ikt}\;,
\ee
for single particle energies.

We now solve the equations of motion and derive the quantisation condition \eqref{eq:quantcon}. Plugging \eqref{eq:matrixel} into the equations of motion we get two coupled differential equations
\be\begin{split}
    ikf_1(x)=&-f_1'(x)+\Theta(-x)(f_1'(0) \pm f_2'(0)) + \Theta\left(x - \tfrac{L}{2}\right)\left(f_1'\left(\tfrac{L}{2}\right) + f_2'\left(\tfrac{L}{2}\right)\right)\\
    &+4L\delta\left(x - \tfrac{L}{2}\right)\sum_{n=1}\lambda_{2n-1}\lim_{y\rightarrow \frac{L}{2}}(f_2^{(2n-1)}(y)-f_1^{(2n-1)}(y))\;,
\end{split}\ee
\be\begin{split}
    ikf_2(x) =& f_2'(x) - \Theta(-x)(f_2'(0) \pm f_1'(0)) - \Theta\left(x - \tfrac{L}{2}\right)\left(f_1'\left(\tfrac{L}{2}\right) + f_2'\left(\tfrac{L}{2}\right)\right)\\
    &+4L\delta\left(x - \tfrac{L}{2}\right)\sum_{n=1}\lambda_{2n-1}\lim_{y\rightarrow L/2}(f_2^{(2n-1)}(y) - f_1^{(2n-1)}(y))\;,
\end{split}\ee
with boundary conditions from \eqref{eq:Phibc}
\be
    f_1(0) = \pm f_2(0) \;,\quad f_1\left(\tfrac{L}{2}\right) = f_2\left(\tfrac{L}{2}\right)\;.
\ee
In the interval $x\in[0,L/2)$ the solution (up to an overall multiplicative constant) is
\be
    f_1(x) = e^{-ikx} \;,\quad f_2(x) = \pm e^{ikx}\;.
\ee
Since the first order differential equations contain a delta function supported at the boundary $x = \frac{L}{2}$, the solution will have a finite discontinuity there. Hence the solutions on the interval $x\in[0,L/2]$ take the form
\be\begin{split}
    &f_1(x) = e^{-ikx} + \Theta\left(x - \tfrac{L}{2}\right)D_1(E_1,E_2)\;,\\
    &f_2(x)= \pm e^{ikx} + \Theta\left(x - \tfrac{L}{2}\right)D_2(E_1,E_2)\;.
\end{split}\ee
The prescription \eqref{eq:deltaprescrip} means the defect terms in the equations of motion don't contain additional delta functions. The equations of motion reduce to a finite term and one proportional to a delta function. The finite piece fixes $D_1$ and $D_2$
\be
    D_1(k)=-D_2(k)=\pm e^{ikL/2}-e^{-ikL/2}\;,
\ee
which is compatible with the boundary conditions \eqref{eq:Phibc}. The term proportional to the delta function gives the quantisation condition
\be\label{eq:quantcon}
    \pm e^{ikL/2} - e^{-ikL/2} = -4iL(\pm e^{ikL/2}+e^{-ikL/2})\sum_{n=1}(-1)^n \lambda_{2n-1}k^{2n-1}\;.
\ee
More explicitly if we have periodic fermions ($+$) we find
\begin{align}
    &D_1(k)=-D_2(k) = 2i\sin\left(\tfrac{kL}{2}\right)\;,\\
    \label{eq:Rquant}&\sin\left(\tfrac{kL}{2}\right) = 4L\cos\left(\tfrac{kL}{2}\right)\sum_{n=1}(-1)^{n-1}\lambda_{2n-1}k^{2n-1}\;,
\end{align}
and if we have anti periodic fermions ($-$)
\begin{align}
    &D_1(k)=-D_2(k) = -2\cos\left(\tfrac{kL}{2}\right)\;,\\
    \label{eq:NSquant}&\cos\left(\tfrac{kL}{2}\right) = 4L\sin\left(\tfrac{kL}{2}\right)\sum_{n=1}(-1)^n\lambda_{2n-1}k^{2n-1}\;.
\end{align}
Note that if we set $\lambda_{2n-1} = 0$ for all $n$ then the quantisation conditions \eqref{eq:Rquant} and \eqref{eq:NSquant} become $\sin\left(\tfrac{kL}{2}\right)=0$ and $\cos\left(\tfrac{kL}{2}\right)=0$ respectively which lead to the one particle spectra $k = \frac{2\pi}{L}n$ and $k = \frac{2\pi}{L}\left(n+\frac{1}{2}\right)$, $n\in\Z$. Hence the quantisation condition \eqref{eq:Rquant} corresponds to the R sector and \eqref{eq:NSquant} corresponds to the NS sector.

In order to obtain the quantisation conditions \eqref{eq:Rquant} and \eqref{eq:NSquant} we inserted the defect at the point $x=\frac{L}{2}$. However if we insert the defect at any other point then we would obtain the same quantisation conditions. We can write the defect operators \eqref{eq:defectops} in terms of the original fermion field $\psi(x,t)$ using \eqref{eq:phipsi}
\be
    H_D^{(2n-1)} = -\frac{2L}{\pi} \psi\left(\tfrac{L}{2},0\right) \lim_{\epsilon\rightarrow0^+}\left( \psi^{(2n-1)}\left(\tfrac{L}{2}+\epsilon,0\right) + \psi^{(2n-1)}\left(\tfrac{L}{2}-\epsilon,0\right)\right)\;,
\ee
where $\psi^{(n)}(x,t) = \partial_x^n\psi(x,t)$. If we want to place the defect at $x=aL$, $a\in[0,1]$ rather than at $x=\frac{L}{2}$ the defect operators are
\be\label{eq:defectaL}
    H_D^{(2n-1)} = -\frac{2L}{\pi} \psi\left(aL,0\right) \lim_{\epsilon\rightarrow0^+}\left( \psi^{(2n-1)}\left(aL+\epsilon,0\right) + \psi^{(2n-1)}\left(aL-\epsilon,0\right)\right)\;.
\ee
We can again fold the line $x\in[0,L]$ at the point where the defect has been placed and introduce the new fields
\be
    \Phi_1(x,t) = \sqrt{\frac{iL}{\pi}}\,\psi(2ax,t) \;,\quad \Phi_2(x,t) = \sqrt{\frac{iL}{\pi}}\,\psi(L-2(1-a)x,t) \;,\quad x\in[0,L/2]\;.
\ee
If the free Hamiltonian \eqref{eq:freehamiltonian} and the defect operators \eqref{eq:defectaL} are written in terms of $\Phi_1$ and $\Phi_2$ we can again find the equations of motion and solving the equations of motion will lead to the same quantisation conditions \eqref{eq:Rquant} and \eqref{eq:NSquant}.

Before describing how to reproduce the quantisation condition \eqref{eq:polynomialba} from the defect quantisation conditions \eqref{eq:Rquant} and \eqref{eq:NSquant} we show that this formalism can be used to formally reproduce the spectrum of the transformed theory under the usual modular transform of the characters i.e. the GGE with only the $I_1$ charge inserted.

\subsection{Inserting $I_1$}
\label{subsec:I1}
We first check that this defect Hamiltonian reproduces the energy spectrum for the transformed GGE with just an additional $I_1$ charge inserted. Adding just an additional $I_1$ charge is equivalent to changing the length of the system. In appendix \ref{app:B} we show via the TBA equations that a change in the system length can naturally be interpreted as a defect. This is just the usual partition functions so the modular transform is known exactly. We will first look at the case
\be
    \Tr_{\text{NS},+}\left(e^{ 2\pi i(\a + \t)I_1}\right) = \Tr_{\text{NS},+}\left(e^{ -\frac{2\pi i}{\a + \t} I_1}\right)\;.
\ee
Consider a rectangular torus with side lengths $R$ and $L$ so the modular parameter is $\t=i\frac{L}{R}$. We also take $\a$ to be pure imaginary and set $\a = i\frac{L\b}{R}$. Hence the energies in the transformed theory satisfy
\begin{align}
    \cos\left( \frac{L(1+\b)}{2\pi}k\right)=0
    \label{eq:I1spec0}
\end{align}
and take the form
\be\label{eq:I1spec}
    \frac{2\pi}{L}\frac{1}{1+\b}\left(n+\frac{1}{2}\right) \;,\quad n\in\N\;.
\ee
The question is: how to reproduce this from the
quantisation condition \eqref{eq:NSquant}?

The initial guess would be that including just $I_1$ ought to correspond to only $\lambda_1$ being non-zero.
If we 
set $\lambda_1 = -\frac{\beta}{8}$ and $\lambda_{2n-1}=0$ for $n>1$, then the 
perturbing operator is 
\be
D = -\frac{\b}{8} H_D^{(1)} = \frac{L\b}{2\pi}\,\psi(L/2)\psi'(L/2)
  = -\frac{L\b}{4\pi} T(L/2)
  \;,
\ee
and the quantisation condition is
\be\label{eq:tothquant}
    \cos\left(\tfrac{kL}{2}\right) = \frac{L\b}{2} \sin\left(\tfrac{kL}{2}\right)k\;.
\ee
This is not the desired relation \eqref{eq:I1spec0}. Instead, 
it matches (46) in \cite{Toth:2006tj}, and we see that the perturbation cannot just be $T(0)$.

However we will see below that we need to add an infinite set of irrelevant operators of higher weight to our Hamiltonian in order to reproduce the required spectrum \eqref{eq:I1spec}. In \cite{Toth:2006tj} these higher weight operators aren't added because they are considering a different physical systems where the spectrum is reproduced without including them.

We now assume $k$ has a power series expansion in $\b$ with constant term $\frac{2\pi}{L}\left(n+\frac{1}{2}\right)$. Solving the quantisation condition perturbatively gives
\be
    k = \frac{2\pi}{L}\left(n+\frac{1}{2}\right)\left(1 - \b + \b^2 - \left(1-\frac{(1+2n)^2\pi^2}{12}\right)\b^3 + \left(1-\frac{(1+2n)^2\pi^2}{3}\right)\b^4 + \dots \right)\;.
\ee
We want to remove the additional terms that start at order $\b^3$ to reproduce \eqref{eq:I1spec}. We do this by adding a term of the form $\b^3k^3$ to the quantisation condition. This corresponds to adding an $H_D^{(3)}$ term to the Hamiltonian. The new quantisation condition is
\be
    \cos\left(\tfrac{kL}{2}\right) = \sin(\pi k) \left(\frac{L\b}{2} k + \frac{1}{3}\left(\frac{L\b}{2}\right)^3 k^3 \right)\;,
\ee
which has solutions
\be
    k=\frac{2\pi}{L}\left(n+\frac{1}{2}\right)\left(1 - \b + \b^2 - \b^3 + \b^4 - \left(1 - \frac{(2n+1)^4\pi^4}{120}\right)\b^5 +\dots\right)\;.
\ee
We can continue to add $\b^{2n-1} H_D^{(2n-1)}$ terms to the Hamiltonian to get the required solution for $k$. If we do this order by order the quantisation condition becomes
\be\label{eq:quantcond}
    \cos\left(\frac{L k}{2}\right) = \sin\left(\frac{L k}{2}\right) \sum_{n=1}^\infty \frac{(-1)^{n-1} 2^{2n}(2^{2n}-1)B_{2n}}{(2n)!}\left(\frac{L \b k}{2}\right)^{2n-1}\;,
\ee
where $B_{2n}$ are the Bernoulli numbers. For $|L\b k|<\pi$ this series converges
\be
	\sum_{n=1}^\infty\frac{(-1)^{n-1}2^{2n}(2^{2n}-1)B_{2n}}{(2n)!}\left(\frac{L \b k}{2}\right)^{2n-1} = \tan\left(\frac{L \b k}{2}\right)\;,
\ee
and the quantisation condition becomes
\be\label{eq:I1quant}
    \cos\left(\frac{L k}{2}\right) = \sin\left(\frac{L k}{2}\right)\tan\left(\frac{L \b k}{2}\right)\;,
\ee
or equivalently 
\be 
 \frac{ \cos\left(\frac{L(1+\b)k}{2}\right)}
{ 
  \cos\left(\frac{L\b k}2 \right)
}=0
\;,
\ee
which is in turn equivalent to the desired relation \eqref{eq:I1spec0} with the desired spectrum
\be\label{eq:I1energies}
    k = \frac{2\pi}{L}\frac{1}{1+\b}\left(n+\frac{1}{2}\right) \;,\quad n\in\Z\;.
\ee
Recall that the series in the quantisation condition \eqref{eq:quantcond} converges for $|L\b k| < \pi$. For a fixed $n$ this requires $\b$ to be in the interval
\be
    \b \in \left(\frac{-1}{|2n+1|+1},\frac{1}{|2n+1|-1}\right)\;.
\ee
But if we want the series to converge for all $n$ then the radius of convergence is 0. This means that our result for the quantised energies \eqref{eq:I1energies} is a formal result.

We now return to the Hamiltonian \eqref{eq:FullH}. In order to get the required quantisation condition, the $\lambda_{2n-1}$ in \eqref{eq:FullH} are
\be
	\lambda_{2n-1} = -\frac{2(2^{2n}-1)B_{2n}}{(2n)!}L^{2n-2}\b^{2n-1}\;.
\ee
Hence the defect Hamiltonian can be written as
\be
	\sum_{n=1}\lambda_{2n-1}H_I^{(2n-1)} = -\frac{i}{L}\left(\Phi_1\left(\tfrac{L}{2},0\right) + \Phi_2\left(\tfrac{L}{2},0\right)\right)\lim_{y\rightarrow \frac{L}{2}}\tanh\left(\tfrac{L\b}{2} \partial_y\right)(\Phi_2(y,0) - \Phi_1(y,0))\;.
\ee
This is an infinite sum of irrelevant operators which means the Hamiltonian may not be well defined.

We can also reproduce the energies for the Ramond sector as well using the quantisation condition \eqref{eq:Rquant}. Again the coefficients in the defect Hamiltonian are
\be
	\lambda_{2n-1} = -\frac{2(2^{2n}-1)B_{2n}}{(2n)!}L^{2n-2}\b^{2n-1}\;,
\ee
so the quantisation condition becomes
\be
    \sin\left(\frac{Lk}{2}\right) = -\cos\left(\frac{Lk}{2}\right)\tan\left(\frac{L \b k}{2}\right) \;\Rightarrow\;
    \frac{\sin\left(\frac{L(1+\b)k}2\right)}
    {\cos\left(\frac{L\b k}{2}\right)}=0
\ee
and hence the one particle energies are
\be
    k = \frac{2\pi}{L}\frac{n}{1+\b} \;,\quad n\in\Z\;.
\ee
These again match the results coming from the modular properties of the the characters.

\subsection{Arbitrary finite collection of $I_{2n-1}$}
We can also use this defect formalism to reproduce the polynomial quantisation condition \eqref{eq:polynomialba} we get from transforming a GGE with multiple charges inserted. Recall that we have the two quantisation conditions for the energy differences $k$, depending on if we are in the NS or R sectors respectively
\be\begin{split}\label{eq:quantcond2}
    &\cos\left(\frac{Lk}{2}\right) = 4L \sin\left(\frac{Lk}{2}\right)\sum_{n=1}(-1)^n \lambda_{2n-1}k^{2n-1}\;,\\
    &\sin\left(\frac{Lk}{2}\right) = 4L \cos\left(\frac{Lk}{2}\right)\sum_{n=1}(-1)^{n-1} \lambda_{2n-1}k^{2n-1}\;.
\end{split}\ee
The required quantisation conditions are polynomials of the form
\be\label{eq:polyquant}
    \sum_{n=1}^N \a_{2n-1}k^{2n-1} + \frac{kL}{2} = m\pi\;,
\ee
where $m\in\Z+\frac{1}{2}$ for the NS sector and $m\in\Z$ for the R sector. In order to get these quantisation conditions we need to tune the $\lambda_{2n-1}$ such that
\be
    L \sum_{n=1}(-1)^n \lambda_{2n-1}k^{2n-1} = \tan\left(\sum_{n=1}^N \alpha_{2n-1}k^{2n-1} \right)\;.
\ee
Formally this can always be done, we just expand the right hand side as a power series in $k$ and find the coefficients $\lambda_{2n-1}$ in terms of the $\a_{2n-1}$. Once we have fixed the $\lambda_{2n-1}$ in terms of the $\a_{2n-1}$ the quantisation conditions \eqref{eq:quantcond2} become
\be\begin{split}
    &\cos\left(\frac{Lk}{2}\right) = \sin\left(\frac{Lk}{2}\right) \tan\left(\sum_{n=1}^N \alpha_{2n-1}k^{2n-1} \right)\;,\\
    &\sin\left(\frac{Lk}{2}\right) = -\cos\left(\frac{Lk}{2}\right)  \tan\left(\sum_{n=1}^N \alpha_{2n-1}k^{2n-1} \right)\;.
\end{split}\ee
These then lead to the required quantisation conditions \eqref{eq:polyquant}. Again, the defect Hamiltonians that give rise to these quantisation conditions contain an infinite number of irrelevant operators and so the question of whether they are well defined is an issue.

\section{Summary and Outlook}

We have reconsidered the expressions for the modular transform of free-fermion GGEs we discussed in \cite{Downing:2021mfw}. We first showed how our original conjectures could be reinterpreted in terms of a defect. This is not an especially new idea - it was possibly first discussed in \cite{Negro} - but does explain the exact results we have in a very appealing physical way.

The same TBA equation \eqref{eq:tbagse} has also arisen previously in \cite{caselle2013quantisation} in a discussion of $T\bar T$ perturbations. 
The context is a bit different - the transmission factor here arises there as the reflection factor for the reflection off a boundary for a theory with a non-trivial bulk scattering. As a result, the spectrum is not the same although formally the integrals are the same. Since the form of the defect transmission factor is exactly the same as the change in the bulk S-matrix for a $T\bar T$ perturbation and since the $T\bar T$ field does approach the $TT$ field as it nears the boundary, it seems very likely that this is more than just a coincidence.

We also generalised the conjectures in \cite{Downing:2021mfw} for the modular transform of a GGE with just the $I_3$ charge to an arbitrary finite number of charges. This leads to very similar expressions over fermion modes which are given by the solutions to higher degree polynomial equations. These, together with the original conjectures, have been proven in \cite{downing2023modular}.

It is straightforward to see that this result also has the interpretation as a defect, and this must be the case for a generic GGE on physical grounds. It is also easy to see directly from the TBA equations - we show this in appendix \ref{app:B} in the case of a massive purely elastic scattering theory.

It is worth noting that this does not yet lead to an in principle closed action of the modular group since the result of a modular transform on a system with a finite number of conserved charges will usually lead to one with an infinite set of charges. Again the sets of infinite charges for which the GGE transform exists and can be defined are also worth investigating further. 

We have also found a formal expression for the defect operator $D(0)$ in the crossed channel so that the Hamiltonian $H_0 + D(0)$ reproduces the spectrum in the defect theory. However the operator used contains an infinite number of irrelevant operators so isn't very natural. 

Having said this, in \cite{LeClair:2021wfd,Ahn:2022pia} theories deformed by an infinite number of irrelevant operators are considered. This irrelevant operators are the $T\bar T$ operator and its higher weight analogues. There it was found that in order for the theories to have a UV completion an infinite number of irrelevant operators must be included in the deformation. This suggests that actually it is natural that our defect Hamiltonian must include an infinite set of irrelevant terms. Perhaps a more natural defect description can be found.

In \cite{hrj:8932} more general modular transformations $\t \rightarrow \hat\t = \frac{a\t+b}{c\t+d}$ were considered. \GW{These allow an expansion as $\tau\to i\infty$, that is for small $q$, and take a similar product form to \eqref{eq:NS+a2ba} and \eqref{eq:NS-a2ba} (although the more general modular transformation formula is not explicitly given in \cite{hrj:8932}, it is contained in the results in that paper).
}
\blank{
The variable $x$ was defined via $\hat\t = \frac{a+ix}{c}$ and the expressions rewritten so that Poisson summation could be performed on $x$. The motivation was to study the behaviour of the generating functions of power partitions near the roots of unity. If this behaviour is understood then the circle method could be applied to obtain approximate expressions for the power partitions themselves.} 
Here we were concerned with finding a physical interpretation for the modular S transform $\t\rightarrow \hat\t = -1/\t$ and so didn't consider \GW{such more} general modular transformations. 
\blank{But one could also study the GGEs as the variables $q_{2n-1} = e^{\a_{2n-1}}$ ($\a_1 = 2\pi i\t$) simultaneously approach roots of unity and derive asymptotic expressions of the coefficients in the expansion in the $q_{2n-1}$'s.
}
\blank{
While the above calculation would be interesting, it does not give the transformation of the GGEs under the full modular group.} 
In this paper the defect was placed horizontally in the original GGE and vertically in the transformed GGE, as in figure \ref{fig:2}, but physically one would expect a general modular transform to correspond to placing the defect at a different angle. Understanding how a general modular transformation acts on the GGE is obviously an interesting question and we are currently investigating it using these ideas of defects placed at an angle.
\section*{Acknowledgments}

We would like to thank A.~Dymarksy, N.~Gromov, S.~Murthy, A.~Sfondrini, R.~Tateo and D.~Zagier for discussions. This work was supported by the EPSRC grant EP/V520019/1.

\appendix
\section{Massless defect TBA}
\label{app:A}

In \cite{Bajnok_2008}, the TBA equations for a single massive fermionic particle in the presence of a defect are presented:
\begin{align}
    E_0 & = - m \int_{-\infty}^\infty \cosh\theta\,
    \log\left(1 + T_+(\tfrac{i\pi}2-\theta) e^{-\epsilon(\theta)}\right)
    \frac{\rd\theta}{2\pi}
    \;,\label{eq:A.1}\\
    \epsilon(\theta)
    & = mL\cosh\theta - \int_{-\infty}^\infty \phi(\theta-\theta')
    \log\left(1 + T_+(\tfrac{i\pi}2-\theta') e^{-\epsilon(\theta')}\right)
    \frac{\rd\theta'}{2\pi}
    \;.\label{eq:A.2}
\end{align}
Here $T_+(\theta)$ is the transmission factor 
for a particle passing through a defect from the right. In this paper we are considering only right-moving particles for which the transmission factor is $T_-(\theta)=T_+(i\pi-\theta)$, so that $T_+(\tfrac{i\pi}2 - \theta) = T_-(\tfrac{i\pi}2 + \theta)$, and free particles for which $\phi(\theta)=0$. This leads to 
\begin{align}
    E_0 & = - m \int_{-\infty}^\infty \cosh\theta\,
    \log\left(1 + T_-(\tfrac{i\pi}2+\theta) e^{-\epsilon(\theta)}\right)
    \frac{\rd\theta}{2\pi}
    \;,\\
    \epsilon(\theta)
    & = mL\cosh\theta 
    \;.
\end{align}
We can now take the massless limit, $m\to 0, \theta \to \infty$, while keeping
$u = m e^\theta/2$ constant, giving the massless defect TBA \eqref{eq:tba2}
\begin{align}
    LE_0 & = -  \int_{0}^\infty \,
    \log\left(1 + T_-(iu) e^{-\epsilon(u)}\right)
    \frac{\rd u}{2\pi}
    \;,\\
    \epsilon(u)
    & = u
    \;.
\end{align}

\blank{
\newpage
In the massive model, we have equation \cite{Downing:2021mfw}(6.9),
\begin{equation}
    RF(R)
= - M\int \cosh\theta \log(1 + \lambda\exp(-R\mathcal{E})) 
    \frac{\rd \theta}{2\pi}
\end{equation}
which we should compare with the expression in \cite{Bajnok_2008} for the ground state energy in the case of constant bulk $S$-matrix and zero phase shift $\phi(\theta)$, and replacing $L$ by $R$ in their formulae (see \cite{Bajnok_2008}(25) and following)
\begin{equation}
    E_0(R) 
 = -M \int \cosh\theta 
    \log(1 + T_+(\frac{i\pi }2 -\theta) \exp(-MR\cosh\theta) 
    \frac{\rd\theta}{2\pi}
\end{equation}
As a consequence we see that the ground state energy is exactly the same as that in a system with a defect with phase shift $T_+(\theta)$
\begin{equation}
    T_+(\frac{i\pi }2 - \theta) 
 =  \lambda\exp(-R\cE + MR \cosh\theta)
\end{equation}

We now consider the system discussed in \cite{Downing:2021mfw} which is a single massless free fermion and a single charge $I_3$.
}

\section{TBA equations for a GGE and a defect}
\label{app:B}

In \cite{Downing:2021mfw}, we considered the TBA equations in the case of a single massive particle in the direct channel (system) where a particle with rapidity $\theta$ has energy $\cal E(\theta)$ and momentum $\cal P(\theta)$ and found the ground state energy in the crossed  channel (system II) is given by the following equations,
\begin{align}
    E_0 
        &= - \int_{-\infty}^\infty \frac{\rd \cal P(\theta)}{\rd \theta}\, \log(1 + e^{-\tilde\epsilon(\theta)})\,\frac{\rd\theta}{2\pi}
    \;,
    \\
    \tilde\epsilon(\theta) 
    &= L{\cal E}(\theta) - \int_{-\infty}^\infty \phi(\theta-\theta') \log(1 + e^{-\tilde\epsilon(\theta')})\,\frac{\rd\theta'}{2\pi}
    \;.
    \label{eq:B.2}
\end{align}
where $\tilde\epsilon(\theta)$ is the pseudo-energy. 
If we take the direct channel system I to be given by a GGE, 
with the energy given by the regular 1-particle energy plus contributions from the conserved charges, and the momentum by the regular 1-particle momentum,
\begin{align}
   {\cal E}(\theta) = m \cosh(\theta) + F(\theta)
    \;,\;\;
    {\cal P}(\theta) = m \sinh(\theta)
\end{align}
then
\eqref{eq:B.2} takes the form
\begin{align}
        \tilde\epsilon(\theta) 
    &= mL\cosh(\theta) + LF(\theta) - \int_{-\infty}^\infty \phi(\theta-\theta') \log(1 + e^{-\epsilon(\theta')})\,\frac{\rd\theta'}{2\pi}
    \;,
    \label{eq:B.2b}
\end{align}
which is precisely of the form
\cite{Bajnok_2008}(25) for the ground state energy in the presence of a defect with transmission factor
%
\begin{align}
    T_+(\theta) = e^{-LF(i\pi/2-\theta)}
    \;.
\end{align}
The TBA equations can be put in the form \eqref{eq:A.1} and \eqref{eq:A.2} with the substitution
\begin{align}
    \tilde\epsilon(\theta) &= \epsilon(\theta) - L F(\theta)
    \;.
\end{align}
If the GGE consists of a set of KdV charges,
$\sum_n \alpha_{2n_1}I_{2n+1}$, where the KdV charges have 1-particle state eigenvalues
$q_{2n+1}\cosh((2n+1)\theta)$,
then
\begin{equation}
    F = \sum_n \alpha_{2n+1}q_{2n+1}\,\cosh((2n+1)\theta)
    \;,\;\;
    T_+(\theta) = e^{-iL\sum_n (-1)^n 
    \alpha_{2n+1}q_{2n+1} \sinh((2n+1)\theta)}
    \;.
\end{equation}
The simplest example, that of a change in the size of the system from $L$ to $L+\beta L$, considered in section \ref{subsec:I1}, can be implemented in the TBA equation \eqref{eq:B.2} by the choice $F =  \beta m\cosh\theta$, and so is equivalent to the insertion of a defect with transmission factor with the universal form
\begin{align}
    T_+(\theta) = e^{i\beta mL\sinh(\theta)}
    = e^{i\beta L P(\theta)}
    \;.
\end{align}
The generalisation from a single particle to a set of particles, following \cite{ZAMOLODCHIKOV1990695}, is straightforward, and so a GGE is in this way always equivalent to a suitable defect for calculating the spectrum in the crossed channel.

\bibliographystyle{JHEP}
\bibliography{Notes}

\providecommand{\href}[2]{#2}\begingroup\raggedright\begin{thebibliography}{10}

\bibitem{Downing:2021mfw}
M.~Downing and G.~M.~T. Watts, \emph{{Free fermions, KdV charges, generalised
  Gibbs ensembles and modular transforms}},
  \href{https://doi.org/10.1007/JHEP06(2022)036}{\emph{JHEP} {\bfseries 06}
  (2022) 036} [\href{https://arxiv.org/abs/2111.13950}{{\ttfamily
  2111.13950}}].

\bibitem{hrj:8932}
D.~Zagier, \emph{{Power partitions and a generalized eta transformation
  property}},
  \href{https://doi.org/10.46298/hrj.2022.8932}{\emph{{Hardy-Ramanujan
  Journal}} {\bfseries {Volume 44 - Special Commemorative volume in honour of
  Srinivasa Ramanujan - 2021}} (2022) }.

\bibitem{downing2023modular}
M.~Downing, \emph{Modular transform of free fermion generalised gibbs ensembles
  and generalised power partitions},
  \href{https://arxiv.org/abs/2310.07601}{{\ttfamily 2310.07601}}.

\bibitem{Toth:2006tj}
G.~Z. Toth, \emph{{A Study of truncation effects in boundary flows of the Ising
  model on the strip}},
  \href{https://doi.org/10.1088/1742-5468/2007/04/P04005}{\emph{J. Stat. Mech.}
  {\bfseries 0704} (2007) P04005}
  [\href{https://arxiv.org/abs/hep-th/0612256}{{\ttfamily hep-th/0612256}}].

\bibitem{Fendley_1992}
P.~Fendley, \emph{Excited-state thermodynamics},
  \href{https://doi.org/10.1016/0550-3213(92)90404-y}{\emph{Nuclear Physics B}
  {\bfseries 374} (1992) 667}
  [\href{https://arxiv.org/abs/hep-th/9109021}{{\ttfamily hep-th/9109021}}].

\bibitem{Negro}
G.~Hern\'andez-Chifflet, S.~Negro and A.~Sfondrini, \emph{{Flow Equations for
  Generalized $T\overline{T}$ Deformations}},
  \href{https://doi.org/10.1103/PhysRevLett.124.200601}{\emph{Phys. Rev. Lett.}
  {\bfseries 124} (2020) 200601}
  [\href{https://arxiv.org/abs/1911.12233}{{\ttfamily 1911.12233}}].

\bibitem{Bajnok_2008}
Z.~Bajnok and Z.~Simon, \emph{Solving topological defects via fusion},
  \href{https://doi.org/10.1016/j.nuclphysb.2008.04.003}{\emph{Nuclear Physics
  B} {\bfseries 802} (2008) 307}
  [\href{https://arxiv.org/abs/0712.4292}{{\ttfamily 0712.4292}}].

\bibitem{caselle2013quantisation}
M.~Caselle, D.~Fioravanti, F.~Gliozzi and R.~Tateo, \emph{Quantisation of the
  effective string with tba},
  \href{https://doi.org/10.1007/JHEP07(2013)071}{\emph{J. High Energ. Phys.}
  {\bfseries 17} (2013) } [\href{https://arxiv.org/abs/1305.1278}{{\ttfamily
  1305.1278}}].

\bibitem{LeClair:2021wfd}
A.~LeClair, \emph{{$T\bar T$ deformation of the Ising model and its ultraviolet
  completion}}, \href{https://doi.org/10.1088/1742-5468/ac2a99}{\emph{J. Stat.
  Mech.} {\bfseries 2111} (2021) 113104}
  [\href{https://arxiv.org/abs/2107.02230}{{\ttfamily 2107.02230}}].

\bibitem{Ahn:2022pia}
C.~Ahn and A.~LeClair, \emph{{On the classification of UV completions of
  integrable $ T\overline{T} $ deformations of CFT}},
  \href{https://doi.org/10.1007/JHEP08(2022)179}{\emph{JHEP} {\bfseries 08}
  (2022) 179} [\href{https://arxiv.org/abs/2205.10905}{{\ttfamily
  2205.10905}}].

\bibitem{ZAMOLODCHIKOV1990695}
A.~Zamolodchikov, \emph{Thermodynamic bethe ansatz in relativistic models:
  Scaling 3-state potts and lee-yang models},
  \href{https://doi.org/https://doi.org/10.1016/0550-3213(90)90333-9}{\emph{Nuclear
  Physics B} {\bfseries 342} (1990) 695}.

\end{thebibliography}\endgroup

\blank{
\section*{Notes}
\parindent 0pt
\parskip 5pt

Defect TBA for ground-state energy
\cite{Bajnok_2008}

Defect TBA equations for purely transmitting - 
Free fermion bilinear defect operators, Castro-Alvaredo and Fring, TBA equations
\cite{Castro_Alvaredo_2003}

$g$-function for defects, M J Martins, 
\cite{Martins_1994}
}

\end{document}